\documentstyle[emulateapj,apjfonts,psfig,natbib]{article}
\def\ra#1#2#3{#1$^{\rm h}$#2$^{\rm m}$#3$^{\rm s}$}
\def\dec#1#2#3{$#1^\circ#2'#3''$}
\def\swift{{\it Swift}}

\def\ociw{1}
\def\prince{2}
\def\hubble{3}
\def\cit{4}
\def\lco{5}
\def\nrao{6}
\def\psu{7}
\def\yale{8}
\def\uh{9}
\def\gsfc{10}
\def\oab{11}

\begin{document}

\title{\large The Discovery of the Optical and Near-IR Afterglows of 
the First \swift\ Gamma-Ray Bursts}

\author{
E.~Berger\altaffilmark{\ociw,}\altaffilmark{\prince,}\altaffilmark{\hubble},
D.~B.~Fox\altaffilmark{\cit},
S.~R.~Kulkarni\altaffilmark{\cit},
W.~Krzeminski\altaffilmark{\lco},
A.~M.~Soderberg\altaffilmark{\cit},
D.~A.~Frail\altaffilmark{\nrao},
D.~N.~Burrows\altaffilmark{\psu},
S.~B.~Cenko\altaffilmark{\cit},
E.~J.~Murphy\altaffilmark{\yale},
P.~A.~Price\altaffilmark{\uh},
A.~Gal-Yam\altaffilmark{\cit,}\altaffilmark{\hubble},
D.-S.~Moon\altaffilmark{\cit},
N.~Gehrels\altaffilmark{\gsfc},
W.~L.~Freedman\altaffilmark{\ociw},
S.~E.~Persson\altaffilmark{\ociw},
S.~Barthelmy\altaffilmark{\gsfc},
J.~E.~Hill\altaffilmark{\psu},
J.~A.~Nousek\altaffilmark{\psu}, 
A.~Moretti\altaffilmark{\oab}
}

\altaffiltext{\ociw}{Observatories of the Carnegie Institution
of Washington, 813 Santa Barbara Street, Pasadena, CA 91101}
 
\altaffiltext{\prince}{Princeton University Observatory,
Peyton Hall, Ivy Lane, Princeton, NJ 08544}
 
\altaffiltext{\hubble}{Hubble Fellow}

\altaffiltext{\cit}{Division of Physics, Mathematics and Astronomy,
105-24, California Institute of Technology, Pasadena, CA 91125}

\altaffiltext{\lco}{Las Campanas Observatory, Carnegie Observatories, 
Casilla 601, La Serena, Chile}

\altaffiltext{\nrao}{National Radio Astronomy Observatory, Socorro,
NM 87801}

\altaffiltext{\psu}{Department of Astronomy and Astrophysics, 
Pennsylvania State University, 525 Davey Laboratory, University 
Park, PA 16802}

\altaffiltext{\yale}{Department of Astronomy, Yale University, P.O.
Box 208101, New Haven, CT 06520-8101}

\altaffiltext{\uh}{Institute for Astronomy, University of Hawaii, 
2680 Woodlawn Drive, Honolulu, HI 96822}

\altaffiltext{\gsfc}{NASA Goddard Space Flight Center, Greenbelt, 
MD 20771}

\altaffiltext{\oab}{INAF -- Osservatorio Astronomico di Brera, 
Via Bianchi 46, 23807 Merate, Italy}

\begin{abstract} 
We present optical and near-infrared searches for afterglow emission
from the first four \swift\ bursts with accurate positions from the
X-ray Telescope (XRT).  Using telescopes at Las Campanas, Keck, and
Palomar observatories we rapidly identified and followed up afterglows
for three of the four bursts.  The burst positions were also observed
with the Very Large Array, but no radio afterglow emission was
detected.  The optical/NIR afterglows are fainter than about $75\%$ of
all afterglows detected to date, with GRB\,050126 being the faintest,
and were identified thanks to accurate and rapid positions from the
XRT and rapid response with $\gtrsim 1$-m telescopes.  This suggests
that the fraction of dust-obscured bursts is small, $\lesssim 10\%$
when combined with afterglows localized by the HETE-2 Soft X-ray
Camera.  The X-ray fluxes are typical of the known population, with
the exception of GRB\,050126 which has the faintest X-ray afterglow to
date (normalized to $\Delta t=10$ hr), and was detected thanks to a
response time of only 130 s after the burst.  Finally, we find that
all three optical/NIR afterglows are located $\lesssim 2$ arcsec away
from the nominal XRT positions, suggesting that the XRT is capable of
delivering highly accurate positions, which will revolutionize
afterglow studies.
\end{abstract}


\section{Introduction}
\label{sec:intro}

The \swift\ $\gamma$-ray satellite \citep{gcg+04}, launched on 2004, 
November 20, holds great promise for our understanding of
$\gamma$-ray bursts (GRBs), as well as their use for cosmological
applications.  This is primarily because of the positional accuracy
and great sensitivity of the Burst Alert Telescope (BAT), and the
on-board X-ray telescope (XRT) and UV/optical telescope (UVOT), which
are capable of providing $\sim 0.3-5$ arcsec positions and detailed
light curves within a few minutes after the burst.  Starting in
mid-December 2004 \swift\ has localized several bursts of which a few
have been followed up with the XRT providing $\sim 8-30$ arcsec error
circles on a timescale of several hours.  The rapidity and accuracy of
these localizations have enabled deep ground-based optical and
near-infrared (NIR) searches.

Here we present a comprehensive investigation (optical, near-IR,
radio) of the first four \swift\ bursts with XRT detections: GRBs
041223, 050117a, 050124 and 050126.  The observations were conducted
at Las Campanas Observatory (LCO), Palomar Observatory, Keck
Observatory and the Very Large Array (VLA).  Even at this early stage,
with the localization timescale and accuracy still an order of
magnitude below the eventual capability of \swift, the combination of
\swift\ and $\gtrsim 1$-m class ground-based telescopes suggests that
the fraction of dust-obscured GRBs is likely low, and the afterglow
recovery rate for \swift\ bursts may approach unity.

\section{Afterglow Identification and Follow-Up of \swift\ Gamma-Ray 
Bursts}
\label{sec:obs}

\subsection{GRB\,041223}

The \swift\ Burst Alert Telescope (BAT) localized this burst on 2004,
December 23.5877 UT to a $7'$ radius error circle
\citep{gcn2898,gcn2909}.  A series of XRT observations was initiated 
on December 23.780 UT, and a fading source was detected at
$\alpha$=\ra{06}{40}{49.2}, $\delta$=\dec{-37}{04}{21.5} (J2000) with
an uncertainty of about $15''$ radius \citep{gcn2901}.  The spectral
energy index was $\beta_x=-1.02\pm 0.13$ and the temporal decay rate
was about $\alpha_x=-1.7\pm 0.2$ ($F_\nu\propto t^\alpha\nu^\beta$)
with a flux of $6.5\times 10^{-12}$ erg cm$^{-2}$ s$^{-1}$ ($0.5-10$
keV) about 6.2 hr after the burst (Table~\ref{tab:swift};
\citealt{bhc+05}).  Following our discovery of the optical transient,
the XRT position was revised to \citep{gcn2910} $\alpha$=
\ra{06}{40}{47.4}, $\delta$=\dec{-37}{04}{22.3} (J2000), within about 
$1"$ of the optical afterglow position.

\begin{figure}
\centerline{\psfig{file=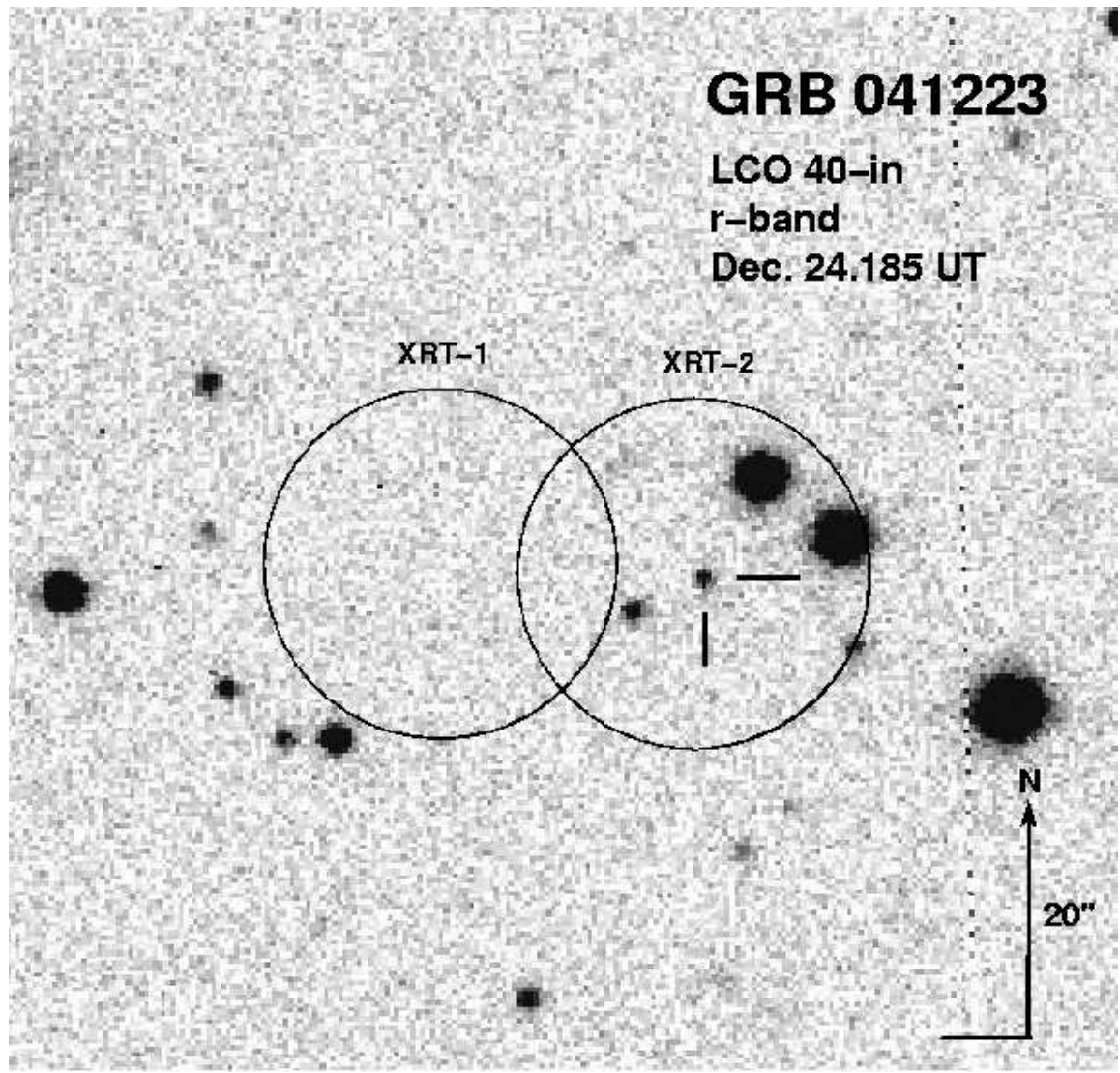,width=6.0in}}
\caption{The field centered on the position of the optical afterglow of 
GRB\,041223 (crosshairs) imaged in the $r$-band with the LCO 40-in
telescope on Dec.~24.185 UT (14.4 hrs after the burst).  Also shown
are the initial (XRT-1) and revised (XRT-2) $15''$ radius error circles
from \swift/XRT.
\label{fig:041223}}
\end{figure}

Ground-based observations commenced on December 24.185 UT (14.4 hours
after the burst) using the Swope 40-in telescope at LCO
\citep{gcn2902}.  We imaged the entire $7'$ radius BAT error circle in
the $r$-band for a total of 20 min.  The data were bias-subtracted,
flat-fielded, and combined using standard IRAF routines.  Astrometry
was performed relative to the USNO-B catalog using 200 stars in common
to the two frames.  The resulting rms positional uncertainty was
$0.15''$.  A stationary source not present in the Digital Sky Survey
(DSS) was detected at $\alpha$=\ra{06}{40}{47.323}, $\delta$=
\dec{-37}{04}{22.77} (J2000) with a magnitude of $r\approx 21\pm 0.15$.  
This position was $7.5''$ outside of the initial XRT error circle, but
only $1''$ from the revised nominal XRT position.  A field centered on
the position of the afterglow of GRB\,041223 is shown in
Figure~\ref{fig:041223}, and the observations are summarized in
Table~\ref{tab:gb}.

Additional observations with the Swope 40-in telescope were obtained
starting on December 25.15 UT in the $r$ and $i$ bands.  A total of 1
hr was obtained in each filter.  A comparison of the first and second
epoch indicated that the afterglow had faded by $1.2$ mag,
corresponding to a decay rate of $\alpha\approx -1.1$.

We subsequently observed the position of the afterglow with the
Low-Resolution Imager and Spectrograph (LRIS; \citealt{occ+95})
mounted on the Keck-I 10-m telescope on 2005, January 8.34 UT.  We
obtained $R$-band observations for a total of 70 min.  The data were
reduced and analyzed in the manner described above.  These
observations reveal a faint source at the position of the afterglow
with $R\approx 24.5\pm 0.3$ mag.  An extrapolation of the afterglow
flux at $t=1.56$ d to the epoch of the LRIS observation suggests that
this object is most likely the afterglow, although any steepening in
the afterglow evolution (e.g., jet break) would mean that the emission
is dominated by the host galaxy.

Late-time observations were obtained with the Near Infra-Red Camera
(NIRC; \citealt{ms94}) mounted on the Keck-I telescope in the
$K_s$-band on 2005, January 25.33 UT.  A total of sixty-two 50-s
images were collected.  The individual images were dark-subtracted,
flat-fielded, and corrected for bad pixels and cosmic rays.  We then
created object masks, which were used to construct improved flat
fields for a second round of reduction.  The data were finally
registered, shifted, and co-added.  Photometry was performed relative
to three 2MASS sources in the field, and no object was detected at the
position of the afterglow to a $3\sigma$ limit of $K_s=22.0$ mag.

Finally, we obtained spectroscopic observations using LRIS with a
$400$-line grating on the red side (dispersion of 1.86 \AA/pix) and a
$600$-line grism on the blue side (dispersion of 0.63 \AA/pix).  Two
2400 s exposures were obtained with a 1.5" slit.  The data were
bias-subtracted and flat-fielded using IRAF.  Rectification and sky
subtraction were performed using the method and software described in
\citet{kel03}.  We detect weak continuum emission, but no obvious
emission lines in the range $\approx 3500-9500$\AA.

\subsection{GRB\,050117a}

This burst was localized by the BAT on 2005, January 17.5365 UT to a
$4'$ radius error circle \citep{gcn2952,gcn2962}.  XRT observations
revealed a fading source at $\alpha$=\ra{23}{53}{53.0}, $\delta$=
\dec{+65}{56}{20} (J2000) with an uncertainty of $15''$ radius
\citep{gcn2955}.  We note that the location of GRB\,050117a less than 
$4^\circ$ away from the Galactic plane results in large extinction,
$E(B-V)=1.75$ mag \citep{sfd98}, which severely hampered optical
searches.

\begin{figure}
\centerline{\psfig{file=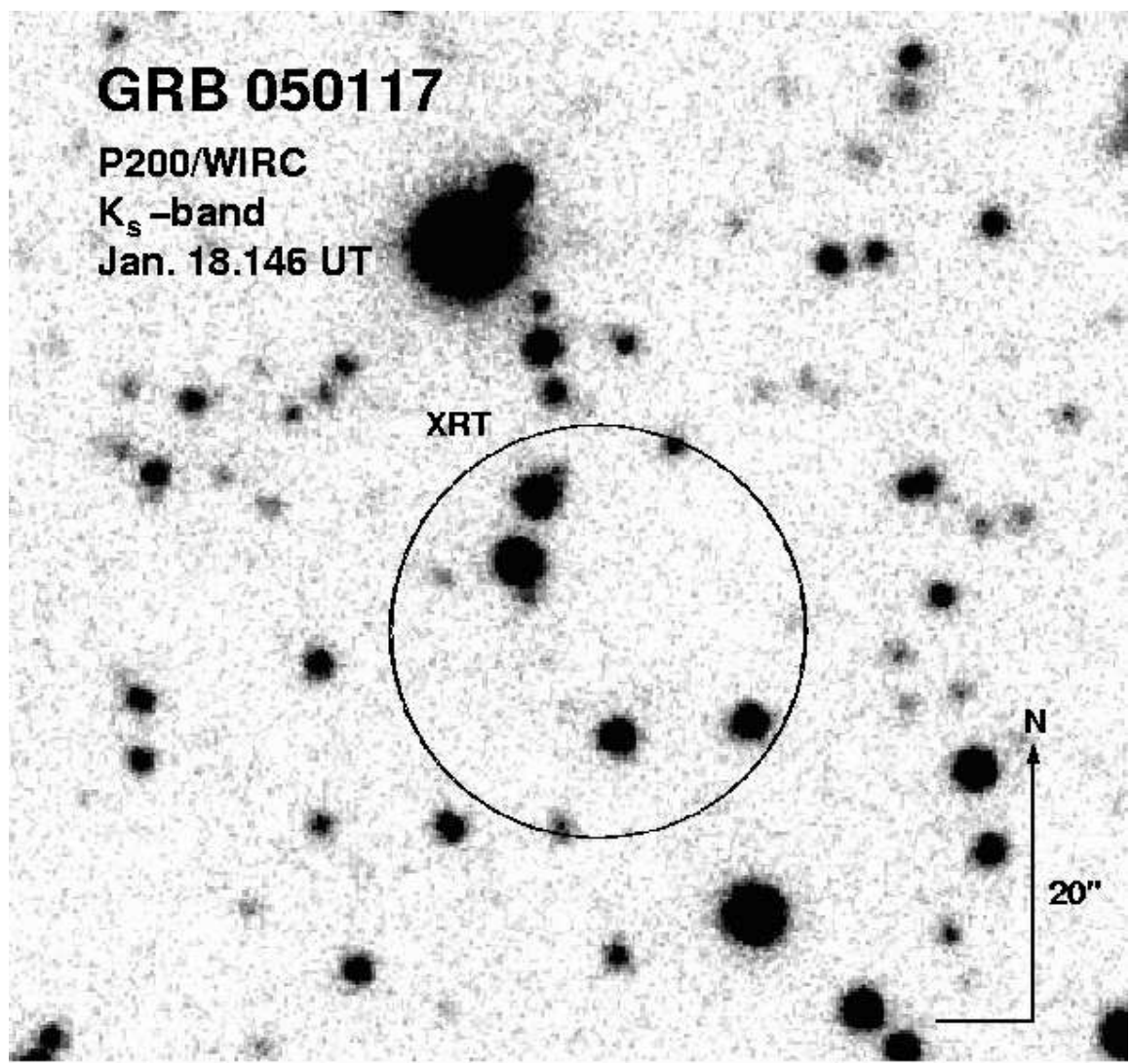,width=6.0in}}
\caption{The field centered on the XRT position of GRB\,050117a imaged 
in the $K_s$-band with WIRC on the Palomar 200-in telescope on
Jan.~18.146 UT (14.6 hrs after the burst).  Also shown is the $15''$
radius XRT error circle.
\label{fig:050117}}
\end{figure}

We observed the XRT position of GRB\,050117a with the Wide Field
Infra-red Camera (WIRC) mounted on the Palomar Hale 200-in telescope
on January 18.146 UT (14.6 hrs after the burst; \citealt{gcn2960}).  A
total of 32 min were obtained in the $K_s$ band.  Several 2MASS and
DSS sources were detected within and near the XRT position.  A field
centered on the XRT error circle of GRB\,050117a is shown in
Figure~\ref{fig:050117}.  At the present no afterglow candidate is
identified.

We observed the field with the VLA\footnotemark\footnotetext{The VLA
is operated by the National Radio Astronomy Observatory, a facility of
the National Science Foundation operated under cooperative agreement
by Associated Universities, Inc.} on 2005, January 19.08 and 24.14 UT
(1.54 and 6.60 days after the burst, respectively) at a frequency of
8.46 GHz \citep{gcn2963,gcn2980}.  No source was detected within the
error circle to a $3\sigma$ limit of 98 (Jan.~19.08) and 84
(Jan.~24.14) $\mu$Jy.

\subsection{GRB\,050124} 

This burst was localized by the BAT on 2005, January 24.4792 UT to a
$6'$ radius error circle \citep{gcn2972,gcn2973}.  An XRT observation
was initiated on January 24.607 UT ($3.1$ hr after the burst), and
ground analysis revealed a source at $\alpha$=\ra{12}{51}{30.4},
$\delta$=\dec{+13}{02}{39.0} (J2000), with an uncertainty of $8''$
\citep{gcn2974}.  The flux of the source was $2\times 10^{-12}$ erg
cm$^{-2}$ s$^{-1}$ ($2-10$ keV).

\begin{figure}
\centerline{\psfig{file=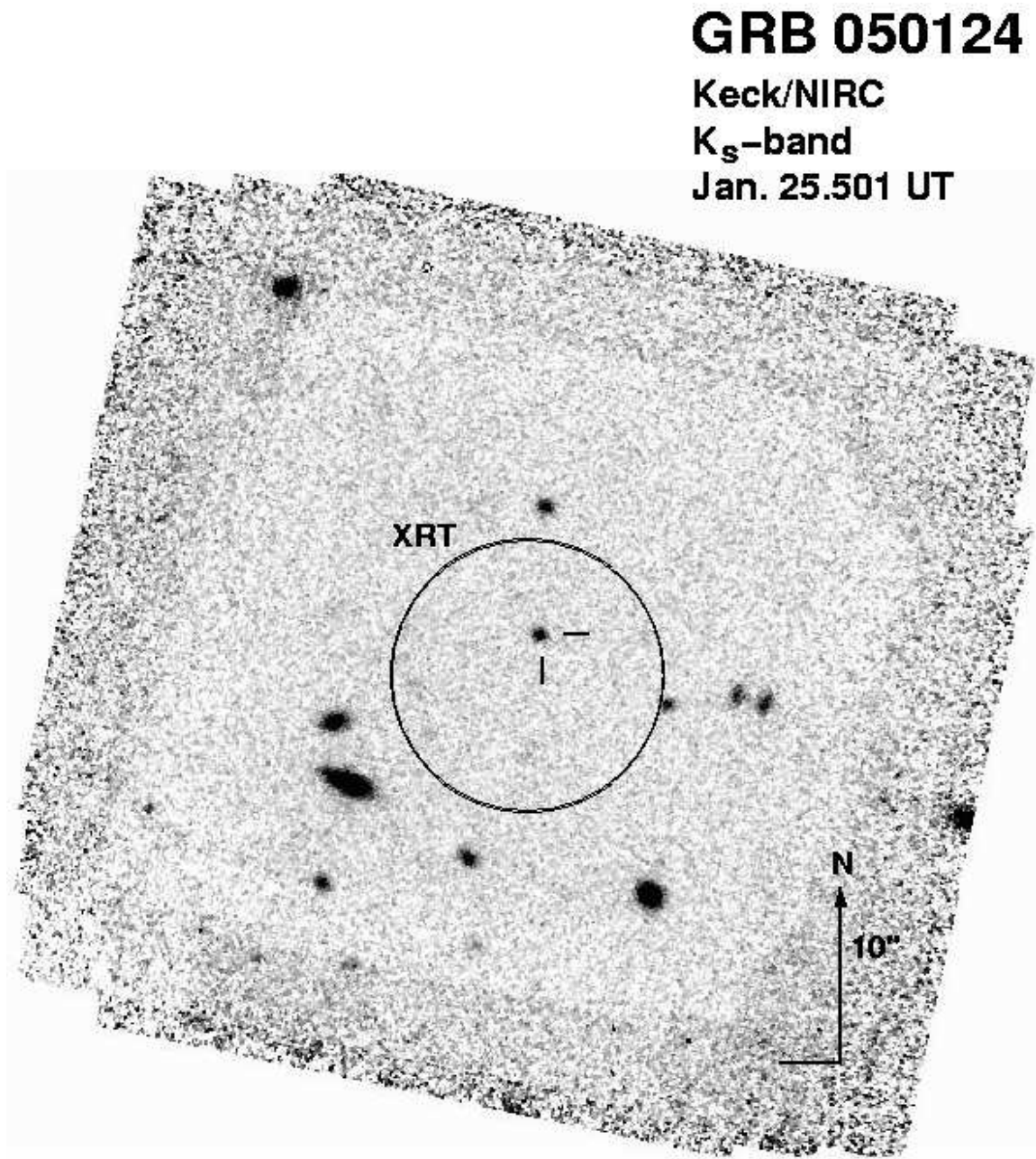,width=6.0in}}
\caption{The field containing the position of the NIR afterglow of 
GRB\,050124 (crosshairs) imaged in the $K_s$-band with NIRC on the
Keck I 10-m telescope on Jan.~25.501 UT (24.5 hrs after the burst).
Also shown is the revised \swift/XRT $8''$ radius error circle.
\label{fig:050124}}
\end{figure}

We observed the XRT $8''$ error circle with NIRC in the $J$ and $K_s$
bands starting on January 25.501 (24.5 hrs after the burst;
\citealt{gcn2978}).  A total of $15$ min were obtained in each band.
Within the XRT error circle we detected a single point source, located
at $\alpha$=\ra{12}{51}{30.35}, $\delta$=\dec{+13}{02}{41.3} (J2000).
The astrometry was performed relative to an image of the field from
the Palomar 60-in telescope with an rms positional uncertainty of
$0.2''$.  The NIR afterglow position is only $2.4''$ away from the
nominal XRT position.  Follow-up observations with NIRC on January
26.471 (47.8 hours after the burst) in the $J$ (13.3 min) and $K_s$
(14.2 min) bands revealed a clear fading of the point source
confirming its identification as the afterglow of GRB\,050124
\citep{gcn2983}.  The brightness of the source was $Ks=19.66\pm 0.06$
mag in the first observation.  The observations are summarized in
Table~\ref{tab:gb} and the first epoch NIRC image is shown in
Figure~\ref{fig:050124}.

Observations were conducted with the VLA at 4.86 and 8.46 GHz on 2005,
January 29.41 UT (4.93 days after the burst; \citealt{gcn3000}).  No
source was detected at the position of the NIR afterglow, or within
the XRT position, to a $3\sigma$ limit of 130 (4.86 GHz) and 100 (8.46
GHz) $\mu$Jy.

\subsection{GRB\,050126}

This burst was localized with the BAT on 2005, January 26.5001 UT to a
$4'$ radius error circle \citep{gcn2987}.  The XRT observation started
129 s after the burst and revealed a source which was localized to a
$30''$ error circle.  Ground analysis based on data from four orbits
resulted in a refined position of $8''$ accuracy \citep{gcn2996}
centered on $\alpha$= \ra{18}{32}{27.0}, $\delta$=\dec{+42}{22}{13.5}
(J2000).

\begin{figure}
\centerline{\psfig{file=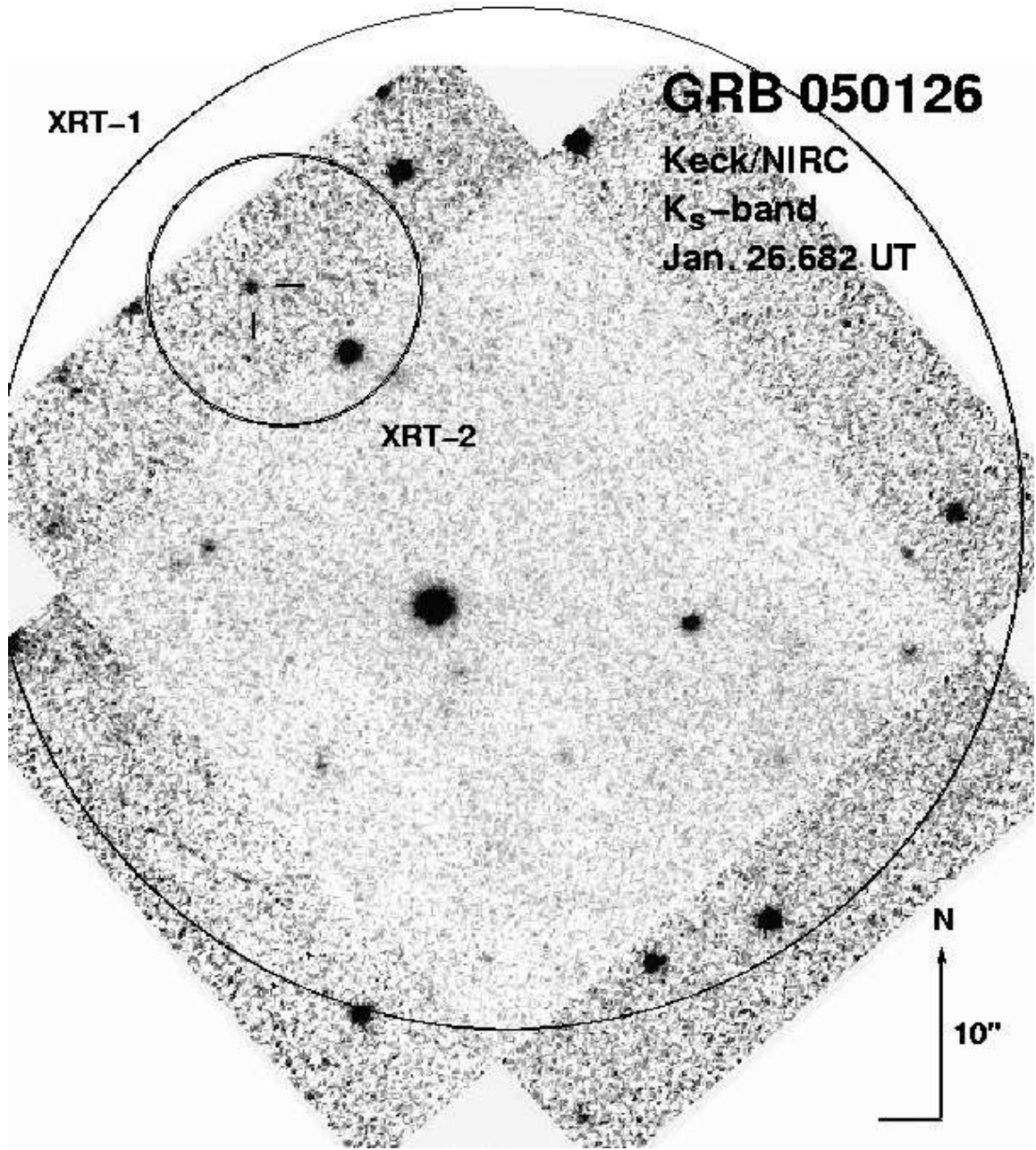,width=6.0in}}
\caption{The field containing the position of the NIR afterglow of 
GRB\,050126 (crosshairs) imaged in the $K_s$-band with NIRC on the
Keck I 10-m telescope on Jan.~26.682 UT (4.4 hrs after the burst).
Also shown are the initial (XRT-1) $30''$ radius \swift/XRT error
circle, and the revised (XRT-2) $8''$ radius error circle.
\label{fig:050126}}
\end{figure}

We observed the XRT $30''$ error circle with NIRC in the $K_s$ band
starting on 2005, January 26.682 UT ($4.4$ hours after the burst;
\citealt{gcn2985,gcn2997}) for a total of $8.3$ min.  The data were
reduced in the manner outlined above, and astrometry was performed
relative to the DSS using six stars in common between the two images.
The resulting rms positional uncertainty was $0.12''$.  Within the
revised XRT error circle we detect one object not visible in the DSS
at $\alpha$=\ra{18}{32}{27.18}, $\delta$=\dec{+42}{22}{13.6} (J2000).
This position is only $2.0''$ away from the nominal XRT position.  The
source has $Ks=19.45\pm 0.17$ mag.  The NIRC image is shown in
Figure~\ref{fig:050126}.  We note that while this is the only
candidate in the XRT error circle, we have been unable to confirm that
the object has actually faded.  If this is in fact not the NIR
afterglow, then this is by far the faintest limit for any afterglow
observed to date (see Figure~\ref{fig:moal} and \citealt{khg+03}).

We observed the position of GRB\,050126 with the VLA at 8.46 GHz on
2005, January 26.67 and $28.59$ UT (4.1 hr and $2.09$ d after the
burst, respectively).  No object was detected at the position of the
NIR candidate or within the revised XRT error circle to a limit of
about 100 $\mu$Jy \citep{gcn2993}.

\section{Afterglow Properties}
\label{sec:ag}

We now provide a simple analysis of the afterglow emission from the
individual bursts.  For GRB\,041223 we combine the data presented in
this paper with measurements in the $J$- and $K$-band from
\citet{bhc+05}.  Correcting for Galactic extinction ($A_R=0.32$,
$A_I=0.23$, $A_J=0.11$, and $A_K=0.04$ mag; \citealt{sfd98}) we find
that the best-fit spectral index using all the available observations
is $\beta=-0.6\pm 0.1$, while the best-fit temporal decay rate is
$\alpha=-1.1\pm 0.1$ In the absence of significant extinction within
the host galaxy, we can use these values, along with the synchrotron
closure relations (e.g.~\citealt{bkb+02}), to determine the value of
the electron distribution power law index, $p$ ($N(\gamma)\propto
\gamma^{-p}$), the geometry of the circumburst environment (ISM or
Wind), and the location of the synchrotron cooling frequency relative
to the optical/NIR band.  Three possibilities exist, namely $\alpha
-3\beta/2=0$ (${\rm ISM_b}$), $\alpha-3\beta/2-1/2=0$ (${\rm ISM_r}$
and ${\rm Wind_r}$), and $\alpha-3\beta/2+1/2=0$ (${\rm Wind_b}$); the
subscript designates whether the cooling frequency is blueward (b) or
redward (r) of the optical/NIR band.  The ${\rm ISM_b}$ closure
relation provides the best result, $-0.2\pm 0.25$, indicating that
$p\approx 2.2\pm 0.2$, and the cooling frequency is located blueward
of the optical/NIR band.  This conclusion is supported by the X-ray
spectral index, $\beta_x\approx -1\pm 0.1$, which indicates $p\approx
2\pm 0.2$.

A comparison of the optical/NIR flux and the X-ray flux, extrapolated
to a common time of 19.6 hr using $\alpha_x=-1.7$, indicates that
$\beta_{ox}\approx -0.65$.  Taken in conjunction with the optical and
X-ray spectral indices, this indicates that the cooling frequency is
$\nu_c\approx 1.1\times 10^{17}$ Hz or about 0.45 keV.  We note,
however, that in the context of this model, the X-ray temporal decay
is expected to be $\alpha_x\approx -1.1\pm 0.1$, which is about
$2.7\sigma$ away from the measured value, $\alpha_x\approx -1.7\pm
0.2$.  The steeper decay may be due to a contribution from inverse
Compton emission \citep{se01}.  We note that \citet{bhc+05} suggest
that the optical/NIR and X-ray afterglows are dominated by two
different physical components.

We perform a similar analysis for GRB\,050124.  Based on the pair of
$J$- and $K_s$-band observations taken on the first and second nights
after the burst, we find a spectral index, $\beta\approx 0.4\pm 0.2$,
and a temporal decay index $\alpha\approx 1.45\pm 0.25$.  These values
satisfy the closure relation for the ${\rm Wind_b}$ case, $-0.35\pm
0.55$, indicating that $p\approx 2.1\pm 0.3$ and the cooling frequency
is located blueward of the NIR bands.  A comparison of the X-ray flux
at 7.1 hours after the burst to the NIR flux, extrapolated to the
epoch of the X-ray observations using the measured value of $\alpha$,
indicates an optical/X-ray spectral index, $\beta_{ox}\approx -0.5$,
in good agreement with the optical/NIR spectral index.  This suggests
that the cooling break is most likely located near the X-ray band.

Finally, for GRB\,050126 we simply note that both the NIR and X-ray
afterglows are fainter than any other afterglow detected to date.  For
the purpose of this comparison we extrapolate the NIR flux to the
optical $R$-band using a typical spectral index of $-0.6$, and the
X-ray flux from 200 s to 10 hr using $\alpha_x\approx -1.3$, which is
typical for X-ray afterglows \citep{bkf03}.

\section{Disucssion and Conclusions}
\label{sec:conc}

\begin{figure}
\centerline{\psfig{file=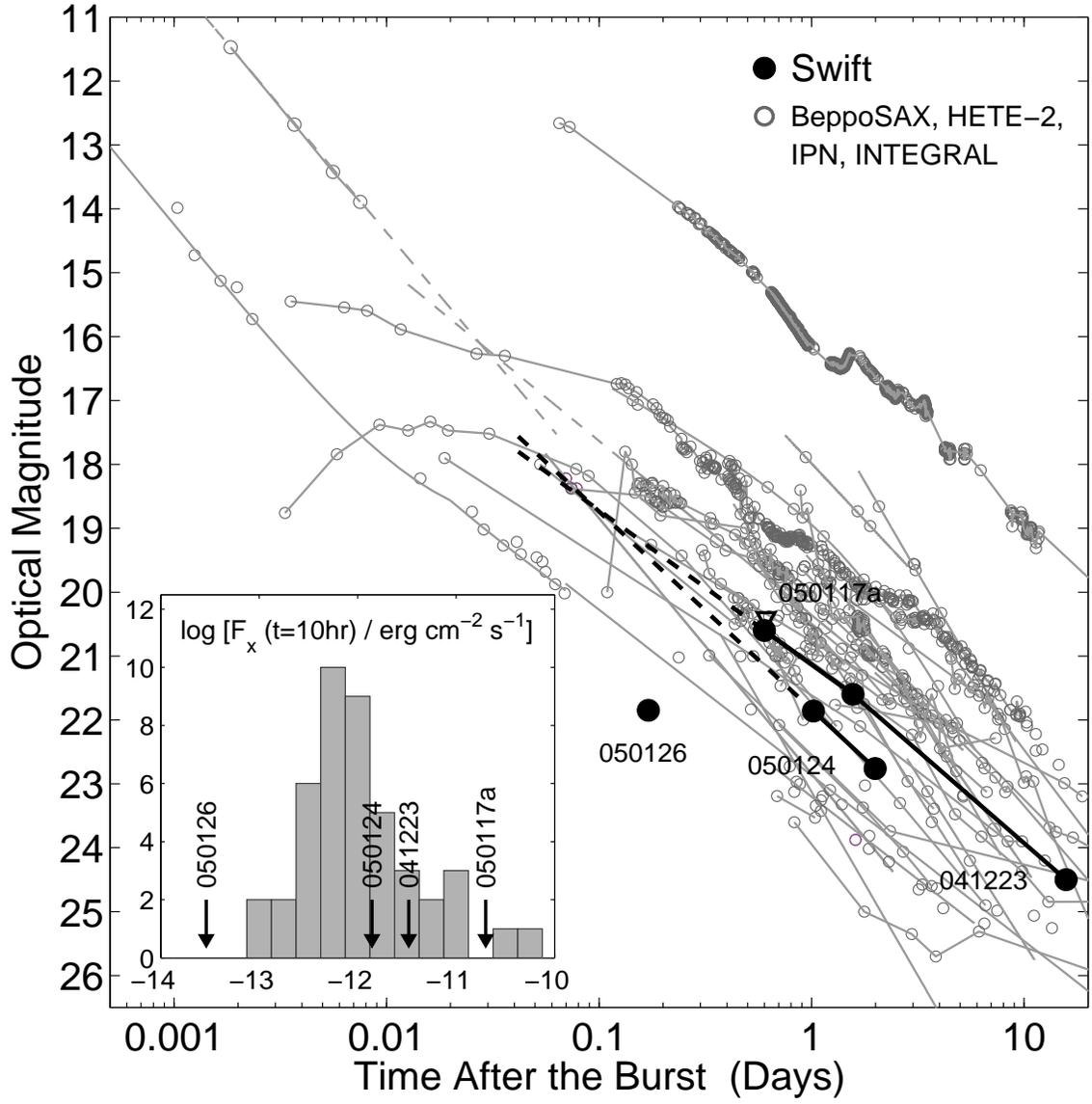,width=6.0in}}
\caption{Optical light curves of the \swift\ bursts discussed in this 
paper (and upper limit on GRB\,050117a) compared to the sample of
afterglows detected and studied in the past seven years.  We
transformed the NIR flux of GRB\,050124 to the $R$-band using the
measured spectral index, and that of GRB\,050126 assuming a typical
index of $\beta=-0.6$.  The \swift\ afterglows are fainter than about
$75\%$ of the known afterglow population.  Their detection was due to
the small error circles from XRT and searches with large telescopes.
The inset shows the distribution of X-ray fluxes at $t=10$ hrs after
the burst for the XRT bursts (using measured temporal decay indices or
assuming the typical $\alpha_x=-1.3$) compared to the sample of
\citet{bkf03}.  Three of the four afterglows are typical of the
general population, but the afterglow of GRB\,050126 is the faintest
detected to date, in agreement with the faintness of the possible NIR
afterglow.  We note that the X-ray emission for GRB\,050117a is
contaminated by the prompt emission and should be considered as an
upper limit.
\label{fig:moal}} 
\end{figure}

One of the main promises of \swift\ is rapid localization and
follow-up with the XRT and UVOT.  The X-ray fluxes from XRT
\citep{bhc+05,hil+05,osb+05,tag+05} are summarized in
\S\ref{sec:obs} and Table~\ref{tab:swift}.  In Figure~\ref{fig:moal}
we plot the X-ray fluxes normalized to 10 hours after the burst in
comparison to the sample of \citet{bkf03}.  We find that three of the
four XRT afterglows are typical of the general population, but the
X-ray afterglow of GRB\,050126 is the faintest detected to date (when
normalized to 10 hrs), and it was only detected thanks to the rapid
response of the XRT.  The fluxes were extrapolated to the common epoch
using the measured temporal decay index or the typical $F_\nu\propto
t^{-1.3}$ \citep{bkf03}.  We note that the X-ray flux for GRB\,050117a
is dominated by the prompt emission and should be considered an upper
limit.

Similarly, we plot the $R$-band magnitudes of the \swift\ afterglows,
measured directly or extrapolated from the NIR (using the measured
spectral indices or a typical $F_\nu\propto\nu^{-0.6}$), in comparison
to a compilation of optical light curves collected in the past seven
years.  We find that the afterglows are fainter than about $75\%$ of
the population on a similar timescale (Figure~\ref{fig:moal}).  In
particular, the possible afterglow of GRB\,050126 is the faintest
detected to date.  The afterglow detections are due to the accurate
positions available from the XRT which allowed us to both identify the
afterglows more readily, and to search the error circles in the NIR
with a large aperture telescope.  We note that the faintness of the
optical afterglows may also explain the non-detections in the radio.

\begin{figure}
\centerline{\psfig{file=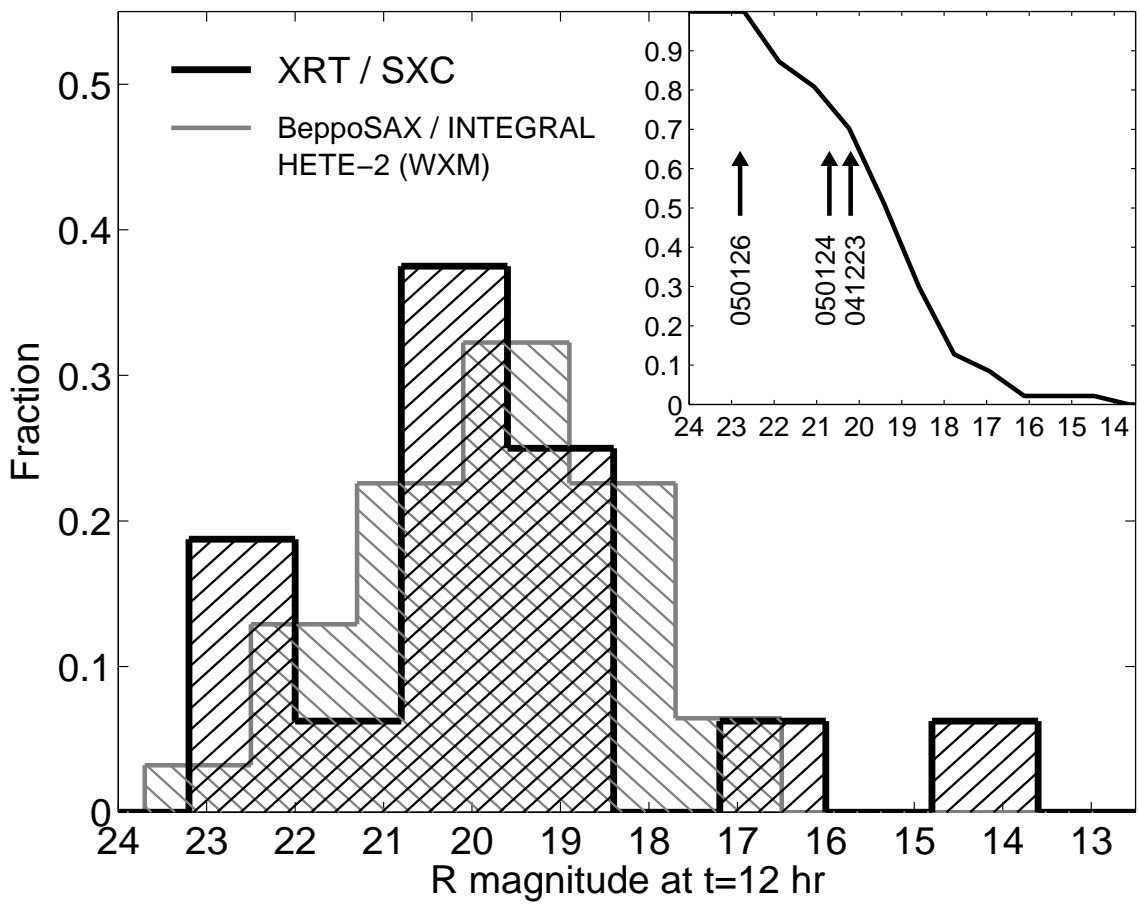,width=6.0in}}
\caption{Distribution of $R$-band magnitudes normalized to 12 hrs after 
the burst for the XRT+SXC sample (black; 16 afterglows) compared to
all other optical afterglows (gray; 31 afterglows).  The afterglow
detection rate for the XRT+SXC sample is about $90\%$ suggesting that
the fraction of dust-obscured (``dark'') GRBs is small.  The inset
shows the cumulative distribution for all afterglows discovered prior
to \swift\ along with the three afterglows discussed in this paper.
The \swift/XRT bursts are fainter than about $75\%$ of all afterglows
localized to date.  In the past, these may have been designated as
dark.
\label{fig:sxcxrt}}
\end{figure}

Several conclusions can already be drawn from this early work.  First,
nearly every XRT localization has resulted in the identification of an
optical or NIR afterglow; the single exception (GRB\,050117a) is likely
due to large Galactic extinction.  The optical/NIR afterglow recovery
rate for XRT (3/4) and the HETE-2 SXC (12/13) is $\gtrsim 90\%$.  The
brightness of the XRT+SXC sample normalized to $t=12$ hr compared to
all other optical afterglows is shown in Figure~\ref{fig:sxcxrt}.  The
afterglows of the XRT bursts appear to be fainter than about $75\%$ of
all afterglows detected prior to \swift, suggesting that past
non-detections were mainly the result of large error regions and/or
shallow searches.  This indicates that the fraction of ``dark''
(dust-obscured) GRBs is low, although we note that two of the XRT
bursts were localized in the NIR and may still be dust-obscured.  If
this trend persists then this bodes well for identification of high
redshift afterglows using the Lyman break technique, since the main
contaminant is dust-obscured bursts.  

Second, in the three cases in which an optical/NIR afterglow was
detected, the offset relative to the nominal XRT position has been
$\lesssim 2$ arcsec, much less than the size of the error circles.
This suggests that in the near future the XRT will be capable of
providing $\sim 2$ arcsec positions.  This will significantly reduce
the delay from localization to identification of the optical/NIR
afterglow in cases when a UVOT sub-arcsecond position is not
available.

We end by noting that the faintness of the optical/NIR afterglows
studied in this paper ($R\approx 18$ mag at $\Delta t=1$ hr;
Figures~\ref{fig:moal} and \ref{fig:sxcxrt}) may make it difficult for
small robotic telescopes to provide long-term follow-up of \swift\
bursts.  However, larger telescopes, while somewhat slower to respond,
will allow both long-term follow-up and detection of the faintest
afterglows, particularly in the NIR.

\acknowledgements We thank the staff at the Las Campanas Observatory,
the Palomar Observatory, the Keck Observatory, and the Very Large
Array.  We also thank Mario Hamuy for generous use of his observing
time and Dan Kelson for help with his sky background subtraction
software.  E.B. is supported by NASA through Hubble Fellowship grant
HST-01171.01 awarded by the Space Telescope Science Institute, which
is operated by AURA, Inc., for NASA under contract NAS 5-26555.
Additional support was provided by NSF and NASA grants.  A.G.
acknowledges support by NASA through Hubble Fellowship grant
HST-HF-01158.01-A awarded by STScI, which is operated by AURA, Inc.,
for NASA, under contract NAS 5-26555.

\clearpage
\begin{deluxetable}{lllllllll}
\tabcolsep0.1in\footnotesize
\tablecolumns{9}
\tabcolsep0.06in\footnotesize
\tablewidth{0pc}
\tablecaption{Prompt Emission and X-Ray Afterglow Properties
\label{tab:swift}}
\tablehead {
\colhead {GRB}        & 
\colhead {$t_0$}      & 
\colhead {$F$}        & 
\colhead {$t_{90}$}   &
\colhead {$\alpha$}   &
\colhead {$t_X$}      &
\colhead {$F_X$}      &
\colhead {$\alpha_X$} &                    
\colhead {$\beta_X$}  \\                   
\colhead {}                         &
\colhead {(UT)}                     &         
\colhead {(erg cm$^{-2}$)}          &         
\colhead {(s)}                      &         
\colhead {}                         &         
\colhead {(s)}                      &        
\colhead {(erg cm$^{-2}$ s$^{-1}$)} &
\colhead {}                         &
\colhead {}
}
\startdata
041223 & 23.5877 & $5\times 10^{-5}$ & 130 & 1.1 & $1.63\times 10^4$ 
& $6.5\times 10^{-12}$ & $-1.72\pm 0.20$ & $-1.02\pm 0.13$ \\
050117a & 17.5365 & $1.7\times 10^{-5}$ & 220 & \nodata & 193 &
$1.8\times 10^{-8}$ $^a$ & \nodata & \nodata \\
050124 & 24.4792 & $2.1\times 10^{-6}$ & 4.1 & 1.5 & $2.54\times 10^4$
& $2.2\times 10^{-12}$ & \nodata & \nodata \\
050126 & 26.5001 & $2.0\times 10^{-6}$ & 26 & 1.3 & 200 & $2.5\times
10^{-11}$ & \nodata & \nodata
\enddata
\tablecomments{Prompt emission and X-ray afterglow properties of the 
four bursts discussed in this paper.  The columns are (left to right):
(i) GRB name, (ii) burst time, (iii) fluence in the $15-350$ keV band,
(iv) burst duration, (v) spectral index, (vi) time of X-ray
observation, (vii) X-ray flux, (viii) X-ray temporal decay index, (ix)
X-ray spectral index.  $^a$ The X-ray flux of GRB\,050117a is dominated
by the prompt emission.}
\end{deluxetable}

\clearpage
\begin{deluxetable}{lllll}
\tabcolsep0.1in\footnotesize
\tablecolumns{5}
\tabcolsep0.1in\footnotesize
\tablewidth{0pc}
\tablecaption{Ground-Based Optical and Near-Infrared Data
\label{tab:gb}}
\tablehead {
\colhead {Date}       &
\colhead {$\Delta t$} &
\colhead {Telescope}  &
\colhead {Filter}     &
\colhead {Magnitude}  \\
\colhead {(UT)}   &
\colhead {(days)} &
\colhead {}       &      
\colhead {}       &      
\colhead {}
}
\startdata
\multicolumn{5}{c}{\bf GRB\,041223} \\\hline
2004, Dec.~24.185 & 0.60 & LCO40 & $r$ & $20.99\pm 0.15$ \\
2004, Dec.~25.204 & 1.62 & LCO40 & $r$ & $22.19\pm 0.14$ \\
2004, Dec.~25.232 & 1.65 & LCO40 & $i$ & $21.82\pm 0.07$ \\
2005, Jan.~8.339  & 15.75 & Keck/LRIS & $R$ & $24.5\pm 0.3$ \\
2005, Jan.~25.333 & 32.75 & Keck/NIRC & $K_s$ & $>22.0$ \\ \hline
\hline\\
\multicolumn{5}{c}{\bf GRB\,050117a} \\\hline
2005, Jan.~18.146 & 0.61 & P200/WIRC & $K_s$ & $>18.5$ \\\hline
\hline\\
\multicolumn{5}{c}{\bf GRB\,050124} \\\hline
2005, Jan.~25.500 & 1.02 & Keck/NIRC & $K_s$ & $19.66\pm 0.06$ \\
2005, Jan.~25.486 & 1.04 & Keck/NIRC & $J$ & $20.90\pm 0.05$ \\
2005, Jan.~26.472 & 1.99 & Keck/NIRC & $K_s$ & $20.63\pm 0.18$ \\
2005, Jan.~26.486 & 2.01 & Keck/NIRC & $J$ & $22.04\pm 0.17$ \\
\hline\\
\multicolumn{5}{c}{\bf GRB\,050126} \\\hline
2005, Jan.~26.682 & 0.18 & Keck/NIRC & $K_s$ & $19.45\pm 0.17$ \\
\enddata
\tablecomments{Ground-based optical and NIR observations of the four 
bursts discussed in this paper.  The columns are (left to right): (i)
UT date of the observation, (ii) time since the burst, (iii)
telescope/instrument, (iv) filter, and (v) observed magnitude (not
corrected for Galactic extinction).  Limits are $3\sigma$.}
\end{deluxetable}

\clearpage

\clearpage

\clearpage

\clearpage

\clearpage

\clearpage

\end{document}